\documentclass[acmsmall]{acmart}

\AtBeginDocument{%
  }

\usepackage{todonotes}
\usepackage{tabularx}
\usepackage{booktabs}
\usepackage{microtype}
\usepackage{cleveref}
\usepackage{makecell}
\usepackage{ragged2e}
\usepackage[table]{xcolor}
\colorlet{lightgray}{gray!15}
\colorlet{darkgray}{gray!45}
\usepackage{pdflscape}

\usepackage{multicol}
\usepackage{multirow}
\usepackage{wrapfig}

\usepackage[T1]{fontenc}
\setcopyright{cc}
\setcctype{by}
\acmDOI{10.1145/3797077}
\acmYear{2026}
\acmJournal{PACMSE}
\acmVolume{3}
\acmNumber{FSE}
\acmArticle{FSE049}
\acmMonth{7}
\acmSubmissionID{fse26maina-p1140-p}
\received{2025-09-11}
\received[accepted]{2025-12-22}

\begin{document}

\title{A Grounded Theory of Debugging in Professional Software Engineering Practice}

\author{Haolin Li}
\orcid{0009-0007-9816-7085}
\affiliation{%
  \institution{University of California at San Diego}
  \city{La Jolla}
  \country{USA}
}
\email{hal180@ucsd.edu}

\author{Michael Coblenz}
\orcid{0000-0002-9369-4069}
\affiliation{%
  \institution{University of California at San Diego}
  \city{La Jolla}
  \country{USA}
}
\email{mcoblenz@ucsd.edu}


\begin{abstract}
Debugging is a central yet complex activity in software engineering. Prior studies have documented debugging strategies and tool usage, but little theory explains how experienced developers reason about bugs in large, real-world codebases. We conducted a qualitative study using a grounded theory approach. We observed seven professional developers and five professional live-coding streamers working on 17 debugging tasks in their own codebases, capturing diverse contexts of debugging. We theorize debugging as a structured, iterative diagnostic process in which programmers update a mental model of the system to guide information gathering. Developers gather information by alternating between navigation and execution strategies, employing forward and backward tracing modes of reasoning and adapting these approaches according to codebase context, complexity, and familiarity. Developers also gather external resources to complement code-based evidence, with their experience enabling them to systematically construct a mental model. We contribute a grounded theory of professional debugging that surfaces the human-centered dimensions of the practice, with implications for tool design and software engineering education.
\end{abstract}



\maketitle

\section{Introduction}
\label{sec:introduction}

Debugging is a central activity in software engineering. Nonetheless, most software engineers were never taught explicitly how to debug~\cite{perscheid2017studying}. Before recommending debugging techniques for use by learners, however, we should first understand what techniques developers find most effective. Previous researchers identified a variety of techniques that developers use to find the causes of bugs, including forward and backward reasoning and hypothesis generation and testing~\cite{Weber2025:Understanding}. However, knowledge of high-level debugging practice has been drawn primarily from surveys and observational studies in which programmers worked in unfamiliar codebases in which they had no long-term stake, or from interviews where participants may give answers to hypothetical scenarios that were different from practice~\cite{Bohme2017:Where,Weber2025:Understanding,arab2025developerschoosedebuggingstrategies,Layman2023:Debugging}.

To deepen our understanding of both high-level and low-level debugging techniques, we observed seven professional software engineers performing debugging work \emph{in situ}, i.e., while doing their regular work on familiar codebases of varying size in which they had a long-term stake. This enabled us to study developers' behavior in contexts in which their long-term understanding of the code would be valuable. To maximize external validity, we did not restrict programming language or application area. We observed work ranging from web applications to operating systems and databases in fields including AI, finance, and medicine. The research question that we focus on is: \emph{How do professional developers debug real-world programming issues?}

To validate and enrich our theory, we analyzed recordings from another five developers who live-streamed their debugging work on six issues. Based on our theoretical framework, we analyzed their use of debugging techniques and time spent at different stages of the debugging process to gain insight into how developers gather information while debugging. 

We adopted a \emph{constructivist grounded theory} methodology~\cite{Charmaz2014:Constructing}, which enables a deep analysis of factors relating to a phenomenon of interest. To our knowledge, we are the first to use grounded theory analysis to study the debugging process. We identified four high-level steps that developers follow in their debugging process: reproduction, mental model development, fix development, and fix validation. Much of our analysis focuses on mental model development, which comprises the bulk of developers' work (\cref{fig:debugging_process}). Although prior work described the use of codebase knowledge as mental models in programming~\cite{LaToza2006:Maintaining}, we approach debugging from a task-specific perspective, where real-time updates to the mental model were the core approach leading up to a fix. 

We show how developers use a \emph{knowledge avoidance} approach to extend their existing models \emph{just barely enough} to fix the bug. This contrasts with earlier advice that developers should construct a thorough understanding of the system by reading manuals, etc.~\cite{AgansDavidJ2006DT9I}. Likewise, although some prior work focuses on scientific hypothesis generation and testing approaches~\cite{zeller2009programs}, we observed an approach that emphasized \emph{uncertainty} and \emph{speculation}. Rather than constructing testable hypotheses, developers often posit vague explanations or theories, refining them iteratively on the basis of information that they gather. This observation may explain a discrepancy between some prior interview-based work, which found that developers \emph{report} using a scientific or hypothesis-based method, and some prior observational work, which found that developers use specific strategies. In fact, \emph{both} are true to some extent, but the scientific process that developers describe may often involve statements that are too vague to be testable as hypotheses, or too many to each be tested.

The approaches and techniques that developers use, such as tracing code at different levels, debuggers, and printf statements, were consistent with prior literature~\cite{Liu2023, arab2025developerschoosedebuggingstrategies}. However, in the real settings we observed, developers more heavily use sociotechnical resources, such as asking co-workers and studying version control history, than has been previously reported. Also, although prior work focuses on strategies individually, we observed that developers combine multiple strategies, using some strategies as subcomponents of other strategies, and use the same strategy with different intentions.

This paper contributes a grounded theory of debugging derived from observations of professional developers' real-world debugging tasks, finding that debugging is an iterative process dominated by mental model development, characterized by knowledge-avoidance strategies, adaptive use of codebase exploration strategies and tracing modes, and external resources to reach the fix.

\section{Method}
\label{sec:methods}

We used a grounded theory approach to construct a theory of debugging from observational data. Our study proceeded in two stages. First, we observed participant sessions (P1--P7) to inductively construct a theoretical understanding of debugging practices (IRB-approved). 
Second, we confirmed and refined the emergent categories by analyzing recordings from professional live-coding developers (S1--S5). This step allowed us to validate the theory across an independent dataset and to further enrich our findings; overall, we studied 12 individuals working on 17 tasks. Although the evidence we obtained from the streamers did not reveal any new theoretical categories, they helped refine our understanding of the categories and relationships we had previously identified. Together, this provides evidence of theoretical saturation.

\subsection{Sampling}

\subsubsection{Participants}
 We recruited experienced software developers who:
\begin{enumerate}
    \item Had at least three years of full-time work experience in software development;
    \item Could share company projects, or a personal project that reflects the programming and debugging skills used in company projects; and
    \item Could bring unresolved debugging task(s), or programming tasks that might involve debugging, to work on during the observation sessions.
\end{enumerate}
To maximize external validity, we did not constrain programming languages, types of debugging tasks, task difficulty, expected duration, or likelihood of resolution. We recruited through direct outreach (e.g., social media posts and developer group events) and industry collaborations (via existing professional networks and partnerships). We recruited seven participants (five men and two women) with an average of 6.5 years of professional experience (range: 3--15 years) and varying levels of leadership experience (range: 0--4 years). Three participants were recruited individually, while four were recruited via three software companies. We encouraged participants to show company projects. Six participants used company projects, and one participant showed a personal project that simulated the techniques used in company projects.

\subsubsection{Streamers}
Our objective of capturing real debugging practices also constrained recruitment. Many companies are reluctant to collaborate due to competing priorities or concerns about sharing sensitive data. Additional challenges included scheduling participants during work hours and ensuring that they brought non-emergent debugging tasks that could realistically be shared in an observation session. Although our high-level theories were established after P6, we extended our dataset by analyzing publicly available live-coding recordings uploaded by professional developers on platforms such as YouTube and Twitch \cite{debugging_episodes, grounded_copilot}. This approach is based on prior work, which found that live-streamed data may reflect a form of pair programming~\cite{Alaboudi2019:Exploratory}. To ensure comparability with our participant sessions, we selected streamers according to the following criteria:
\begin{enumerate}
    \item The recording occurred within one year of the data collection period (August 2025);
    \item The streamer explicitly self-identified as a professional software engineer (e.g., by describing years of professional or open-source experience) and works in English;
    \item The recordings were publicly accessible;
    \item The streamer verbalized their thought process while primarily focusing on coding, approximating a think-aloud protocol;
    \item The recording captured a full live session rather than edited highlights;
    \item The session included at least one non-trivial debugging episode ($\ge$10 minutes, matching the shortest observed participant task).
\end{enumerate}
We excluded sessions centered on block-based environments, low-level assembly, graphics/game debugging, or ``vibe coding'' sessions dominated by AI-generated debugging attempts ~\cite{ray_vibe_coding_2025} because we were interested in studying techniques experts use when debugging general-purpose code. For the streamers, debugging moments are identified by either a debugging ticket, a clear defect, or unexpected behavior, followed by the participant attempting to fix it. The end of the debugging episode is identified as either the streamer's announcement that they gave up or had to continue another time, or the issue and related issues triggered in the process of debugging are fixed, where the developer showed a clear indication of moving on to another task. Although we expected to observe bugs related to run time behavior, we included one session involving a difficult syntax error because it required a related set of problem-solving skills. We recruited a total of five streamers that satisfied the above criteria.

\subsection{Study Procedure for Participant Observations}
All sessions were conducted via video conferencing platforms by the first author. We required all participants to screen share and encouraged them to turn on their camera. For most participants, we obtained consent for both audio and screen recordings to further analyze. Due to intellectual property concerns, one participant allowed us to audio record only. Some companies maintained ownership of the recordings and shared access with the researchers for data analysis. We conducted a pilot study session, during which we estimated appropriate times for each step, after which we established the protocol below. The study protocol script is presented in the supplement.

\textit{Setup (15 mins).} The first author obtained informed consent and asked participants to provide background information about the codebase, their role in the project, and the specific debugging tasks they were addressing. We asked participants to treat the researcher as a new software developer of the team.

\textit{Development (30+ mins).} We asked developers to think aloud~\cite{lewis1982using}, treating the experimenter as a pair programming partner. The researcher asked clarifying questions about the code and the reasoning behind participants' decisions. While the think-aloud protocol may have diverted participant attention and increased the time spent on each task, none of the participants reported that their performance was significantly affected. The researcher took notes on key observations and prepared for follow-up questions during the semi-structured interview. Participants were encouraged to show more than one task if time allowed.

\textit{Follow-up Interview (10 mins).} We conducted semi-structured interviews at the end of the session. We asked participants to discuss their high-level focus and use of debugging methods. Some of our semi-structured interview questions were informed by NASA-TLX~\cite{NASA-TLX}, exploring cognitive experiences when handling debugging tasks.

\subsection{Data Analysis}

We adopted a constructivist grounded theory approach~\cite{Charmaz2014:Constructing,stol2016grounded} for analysis. While we drew insights from existing debugging studies to contextualize our understanding, we employed open coding and allowed theories to emerge inductively from the data.

During each observation session, the first author took field notes to capture immediate impressions and guide the follow-up interviews. After each session, we transcribed the audio recordings.
We added annotations of important observations or developer actions using the corresponding recordings to provide more context. The first author then performed initial coding on the transcripts and discussed the codes with the second author, referencing the recordings while coding. Rather than conducting fine-grained coding of screen-captured actions such as search and navigation, we prioritized coding the verbal transcripts to capture programmers' reasoning process. Body language (as recommended by Charmaz~\cite{Charmaz2014:Constructing}) and screen recordings helped situate the verbalization and relate observable interactions to the underlying debugging logic. 

\begin{table*} 
    \caption{Analysis sample from P7TA (anonymized). Clusters are represented as \textit{category: subcategory (other terminologies)} to show hierarchical coding.}
    \label{tab:gt_sample}
    \renewcommand*{\arraystretch}{1.2}
    \small
    \begin{tabularx}{\linewidth}{p{0.3\linewidth}p{0.28\linewidth}p{0.34\linewidth}}
    \toprule
    \textbf{Excerpt} & \textbf{Initial Coding} & \textbf{Focused coding}\\
    \midrule
    \rowcolor{lightgray}
    {[Uses a breakpoint to step into a function]}
        & inspects the code by following the calls using a breakpoint 
        & codebase debugging: execution (run-time analysis/interactive code comprehension): breakpoint; tracing: forward \\
    Yeah, this is the one that we need.
        & identifies a function relevant to the issue
        & relevant information \\
    \rowcolor{lightgray}
     {[Reads the name of a function called in the function]}, {[Reads variable names in the function]}
        & reads function quickly and line-by-line, utters the function name and tracks key variables
        & codebase debugging: navigation (static code comprehension); tracing: forward \\
    Yeah, I don't know what is going on here.
        & expresses uncertainty towards the function
        & feelings: uncertain (confusion) \\
    \rowcolor{lightgray}
    So this is one of the cases, for example, where I'll use AI to explain what they want to do.
        & identifies a case of using AI to understand a function
        & external resources: AI; planning \\ \hline
    \multicolumn{3}{X}{Memo: P7 demonstrates a cycle of using a debugging technique (breakpoint) to trace to a function identified as relevant to the issue. This narrows the debugging space down to the function level. However, upon a quick function inspection, the participant expressed uncertainty toward the function and turned to use AI as a support for debugging, relying on it as an external source. Interestingly, P7 remarks ``this is one of the cases'' of using AI, showing that the choice of using AI is intentional and dependent on the scenario.} \\
    \bottomrule
    \end{tabularx}
\end{table*}

Our analysis initially focused on identifying patterns in how and why participants used particular debugging strategies and methods. Following grounded theory procedure~\cite{Charmaz2014:Constructing}, we analyzed the data continuously rather than after collecting all data, first through initial open coding. By the fourth participant, recurring patterns and clusters of codes began to emerge across different participants and debugging scenarios. These evolving categories led us to refine our focus during subsequent observations and return to earlier transcripts and screen recordings to conduct focused coding around these behaviors and mapping connections by building versions of the debugging process graph. In later sessions, we asked participants to explain how they locate faulty code quickly. We also wrote analytic memos to capture emerging ideas, document coding decisions, and trace the development of core categories.  This procedure enabled us to identify the categories that structured our findings. An example of our analytic process is shown in Table~\ref{tab:gt_sample}. With the categories emerging, we did not perform initial coding on the streamer data and instead only coded selectively.

\section{Limitations}
Our sample of participants and tasks was limited by our requirement that participants be able to show us their code and contexts, limiting the external validity of our study. Also, because we required arranging sessions in advance, we could not observe debugging in emergent settings, and some observations started after the beginning of debugging or ended before the bug was resolved. Supplementing our data with livestreamed sessions mitigates the latter limitation to some extent.

Because our study relied on participants' verbalization and observable actions, we may not have captured all aspects of their reasoning. Think-aloud protocols can slow task completion, increase fatigue, and possibly shift participants' strategies; additionally, verbalization often emphasizes immediate actions over higher-level goals and plans~\cite{van_den_Haak01092003, Ericsson_Simon_1993}. We mitigated these risks in two ways. First, our script framed the study as ``debug as you would with a new team member watching,'' aligning the protocol with lightweight pair programming, which participants described as similar to their normal workflow (e.g., P1: ``I felt that I was debugging with just [a] coworker here (...) it's something that I'm used to when we collaborate.''). Second, we conducted short follow-up interviews to elicit participants' high-level reasoning, goals, and strategies. Despite these mitigations, participants may still have reasoned silently, and we did not get a chance to prompt think-aloud for streamers. This poses a risk to construct validity, as the data may not fully reflect underlying cognitive processes and required interpretation of actions.

Constructivist grounded theory is inherently interpretive~\cite{Charmaz2014:Constructing}: researchers construct meaning from participants' accounts rather than treating data as neutral facts. Backgrounds and assumptions of the research team may have shaped our data collection and our analysis. Both authors received systematic training in computer science and software engineering, each with dual background in research and industry. This background enabled us to follow participants' reasoning and to interpret their verbalization and screen actions with technical accuracy. At the same time, our analytical lens might be influenced by our training, existing literature on debugging, and familiarity with traditional software engineering practices. We mitigate these risks through common strategies of open coding, memo writing, iterative discussion, and revisiting transcripts to revise interpretations.

\section{Theoretical Results}
\label{sec:theory}

\begin{table}
\centering
\caption{Participant Codebase Information. Field is identified by the authors; other fields are self-reported.}
\label{tab:participant_codebase_info}
\small
\begin{tabularx}{\linewidth}{lXlll}
\toprule
\textbf{ID} & \textbf{\shortstack[l]{Languages \& Frameworks}} & \textbf{Field} & \textbf{\shortstack[l]{Lines of Code}} & \textbf{\shortstack[l]{Codebase\\Familiarity}} \\ \hline
\rowcolor{lightgray}
P1 & C & Kernel source code development & >1m & <5\% \\

P2 & React, JavaScript, Redux & Banking web app& 100k--1m & >=50\% \\

\rowcolor{lightgray}
P3 & JavaScript, Java & Low-code development web app & 10k--100k & <20\% \\

P4 & JavaScript, React & Data dashboard web app & <=10k & <20\% \\

\rowcolor{lightgray}
P5 & \shortstack[l]{TypeScript, React, Python} & Productivity web app& 10k--100k & >=50\% \\

P6 & \shortstack[l]{JavaScript, React, Vue} & Healthcare web app& <=10k & >=50\%\\

\rowcolor{lightgray}
P7 & \shortstack[l]{TypeScript, Node.js} & AI in healthcare web app& 100k--1m & <20\% \\
\bottomrule
\end{tabularx}
\end{table}

\begin{table}
\centering
\caption{Streamer Codebase Information (Inferred by the Authors).}
\label{tab:streamer_codebase_info}
\small
\begin{tabularx}{0.80\linewidth}{lXll}
\toprule
\textbf{ID} & \textbf{\shortstack[l]{Languages \& Frameworks}} & \textbf{Field} & \textbf{\shortstack[l]{Lines of Code}} \\ \hline
\rowcolor{lightgray}
S1 & Rust, CDK, AWS Lambda & Web analytics & <=10k \\

S2 & C, GTK & Chat client & 100k--1m \\

\rowcolor{lightgray}
S3 & C & Run-time developer tool & <=10k \\

S4 & Python & Browser extension & Not Available \\

\rowcolor{lightgray}
S5 & C\# & OS developer tool & 100k--1m \\
\bottomrule
\end{tabularx}
\end{table}

\begin{table*}
\centering
\caption{Combined Participant and Streamer Task Information. Task time is considered from the start of initial bug reproduction to the end of the last step in the debugging process, excluding breaks.}
\label{tab:combined_task_info}
\small
\begin{tabularx}{0.7\linewidth}{llXl}
\toprule
\textbf{Task} & \textbf{Time} & \textbf{Task Summary} & \textbf{\shortstack{Resolved}} \\ \hline

\rowcolor{lightgray}
P1TA & 2 h 25 min & Deadlock issue & Y \\

P2TA & 18 min & UI modal reset issue & Y \\

\rowcolor{lightgray}
P2TB & 12 min & Page crash & Y \\

P3TA & 36 min & Web accessibility issue & Y \\

\rowcolor{lightgray}
P4TA & 32 min & Bad performance on page load & N \\

P5TA & 10 min & Dropdown not resetting & Y \\

\rowcolor{lightgray}
P5TB & 25 min & Auto populate issue & Y \\

P5TC & 25 min & Log out behavior & Y \\

\rowcolor{lightgray}
P6TA & 45 min & Information parsing issue & Y \\

P6TB & 40 min & Save file feature & Y \\

\rowcolor{lightgray}
P7TA & 1 h 20 min & Rich text parsing issue & Y \\

S1TA & 11 min & Syntax error over data extraction & Y \\

\rowcolor{lightgray}
S2TA & 50 min & Fix connection to server issue & Y \\

S3TA & 26 min & Unexpected benchmark behaviors & N \\

\rowcolor{lightgray}
S4TA & 10 min & Incorrect SHA transformations & Y \\

S4TB & 27 min & Incorrect SHA content & N \\

\rowcolor{lightgray}
S5TA & 1 h 54 min & Icon triggering issue & N \\
\bottomrule
\end{tabularx}
\end{table*}

We recruited seven participants (P1-P7) working on 11 debugging tasks totaling 7.8 hours of debugging. To validate our findings, we also analyzed data from five streamers (S1-S5) who worked on six debugging tasks over 3.1 hours. Overall, the dataset comprises 17 debugging tasks across 11 hours within a broad range of debugging contexts and challenges. \Cref{tab:streamer_codebase_info,tab:combined_task_info,tab:participant_codebase_info}  summarize the participant and streamer codebase and task information. The task IDs are unique and combine a task identifier with the developer ID (e.g., P2TA represents the first task (``A'') done by P2).

\subsection{Overview}
\label{sec:theory_overview}

\begin{figure}
    \centering
    \includegraphics[width=0.95\linewidth]{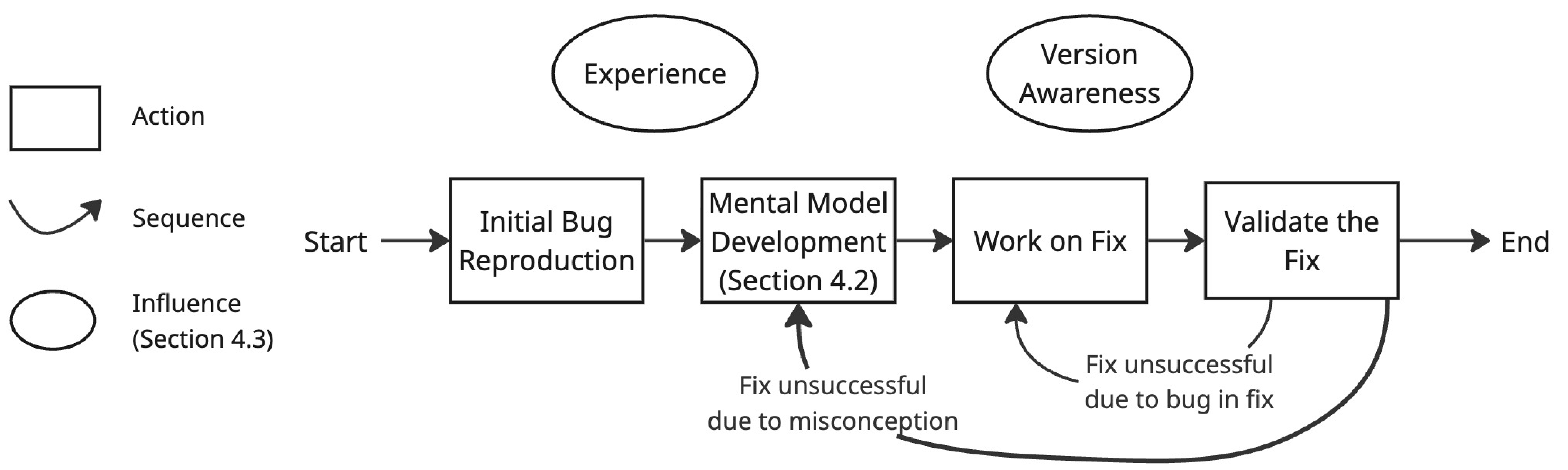}
    \caption{The Debugging Process Overview. Developers cycle through debugging, implementing a fix, and validating the fix. The influence of experience and version awareness guides the entire debugging process. Mental model development process is detailed in Fig. \ref{fig:mental_model} in Section~\ref{sec:mental_model_update}.}
    \label{fig:debugging_process}
\end{figure}

The debugging process we identified comprises four steps. P5 explained, 
\begin{quote}
``I usually have certain steps I take. The first one is triaging (…). Next step is finding the effective code (…). Next part is testing (…). My final step is always to ask myself, is there anything I can do to ensure this does not happen again?''
\end{quote}
That is, debugging is not only a set of strategies, such as inserting \texttt{printf}-style statements, but a process that spans reproduction, verification, and prevention. We observed these steps in both participants and streamers. Our theory, which summarizes this process, is as follows (\cref{fig:debugging_process}):

\begin{enumerate}
    \item \label{step:reproduce} \textbf{Initial bug reproduction.} Developers execute a sequence of steps --- often included in a bug report --- to observe the system producing the incorrect behavior that was to be fixed. The term ``initial'' emphasizes that reproduction often occurs repeatedly throughout debugging. For example, after initial reproduction, P1 reproduced the bug in an isolated example, and P6 reproduced the bug in a local environment. Subsequent reproductions are part of \textit{mental model development} (\ref{step:debugging}) rather than this initial step.
    \item \label{step:debugging} \textbf{Mental model development.} Developers gather evidence to construct and refine a mental model of the bug. This step, detailed in Section~\ref{sec:mental_model_update}, typically consumes the most time.
    \item \textbf{Work on fix.} \label{step:fix} After the mental model helps the participant reach a level of certainty, the developer constructs a candidate to fix the bug.
    \item \textbf{Validate the fix.} Developers attempt to reproduce the bug (\ref{step:reproduce}) to verify whether it was fixed. They also assess the fix in terms of risk, style, and design. If the fix is acceptable, debugging ends. If not, developers return to step~\ref{step:debugging} or \ref{step:fix} to construct a new fix.
\end{enumerate}


The high-level process illustrates the handoffs between stages, defines each stage, and highlights practical lessons for systematic debugging. However, exceptions to this high-level process can occur. For example, S1TA involved a non-trivial syntax error, but bug reproduction and fix validation were trivial because the syntax error could be detected and validated immediately upon editing the code. Second, debugging can be unpredictable when developers are uncertain about the source of the issue or whether it has already been resolved. For instance, P1TA required multiple reproductions beyond the initial bug, including narrowing down the version and isolating a minimal failing case. After an extended effort, P1TA discovered that the bug had already been fixed in the latest patch, skipping the fix construction step. Lastly, in some cases, the process was incomplete because participants did not resolve the task within the study period (P4TA, S3TA, S4TB, S5TA). Despite such exceptions, we observe consistent patterns across most debugging tasks. In support of the grounded theory analysis, Section~\ref{sec:time_analysis} presents the time breakdown of the process. 

Participants' time was dominated by step \ref{step:debugging}, so we devote much of the rest of the paper to this process (Section~\ref{sec:mental_model_update}).
Section~\ref{sec:influences} discusses how experience and version awareness influence the debugging process. We omit further discussion of bug reproduction and fix development and validation, as our data do not focus on these areas.

\begin{figure}[tb]
    \centering
    \includegraphics[width=.95\linewidth]{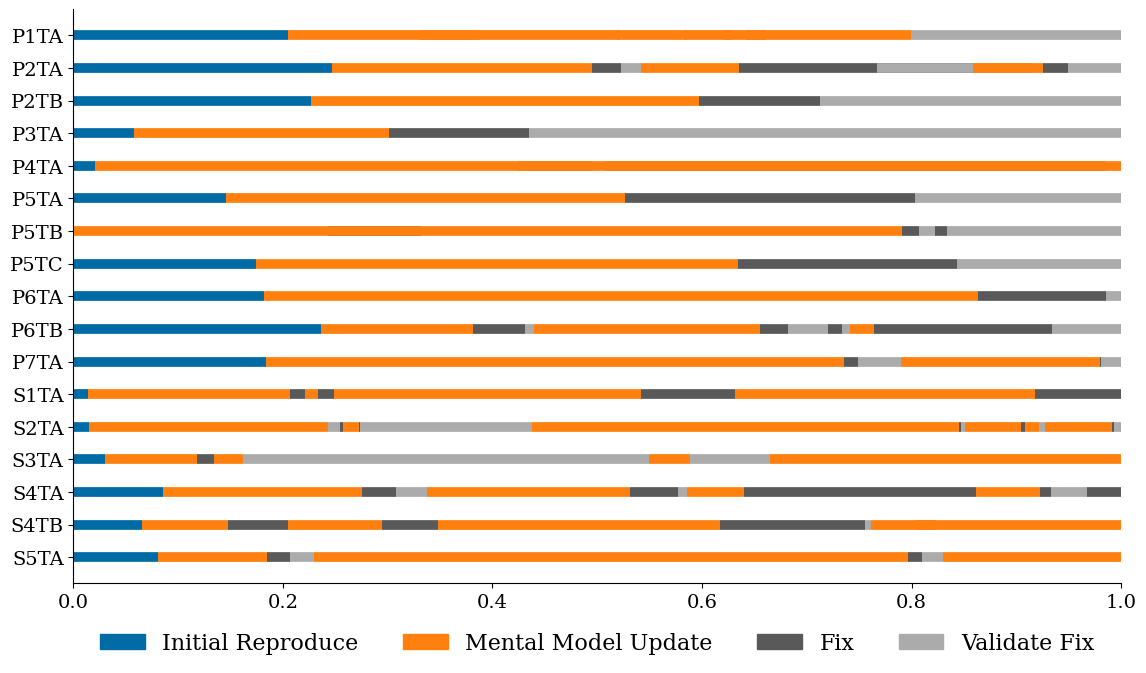}
    \caption{Debugging Steps Over Normalized Time.}
    \label{fig:time_sequence}
\end{figure}

\label{sec:time_analysis}
\Cref{fig:time_sequence} shows how our participants spent their time in the high-level steps of the debugging process (Section~\ref{sec:theory_overview}). On average, participants spent 12\% of their time reproducing the issue, 57\% of their time updating the mental model, 13\% of their time fixing the issue, and 18\% of their time validating the fix. The majority of time was spent on updating the mental model, with substantially less time spent on actual fixing and validation. 

On average, each task went through 1.65 fix --- validate fix cycles. In 9 out of the 17 tasks, a participant returned to the \emph{mental model update} step after testing an initial fix. The four sessions that were not finished at the end of the observation ended in mental model update.

Streamers made significantly more progress transitions than participants (median = 9 vs. 4 process count, $p = 0.014$) as determined by a Mann--Whitney U test. An explanation is that participants might be more careful in showing their work as they are less familiar with being observed and want to show their best performance in a study setting. P5, for example, committed code after task A and said ``I am not always this diligent when it comes to personal projects,'' showing potential for behavioral changes towards being more careful and clean in working on the task. On the other hand, streamers might be faster-paced in their debugging process to keep the watchers involved, and they are likely more accustomed to talking over the development process.


\subsection{Mental Model Development}
\label{sec:mental_model_update}

\begin{figure}
    \centering
    \includegraphics[width=0.95\linewidth]{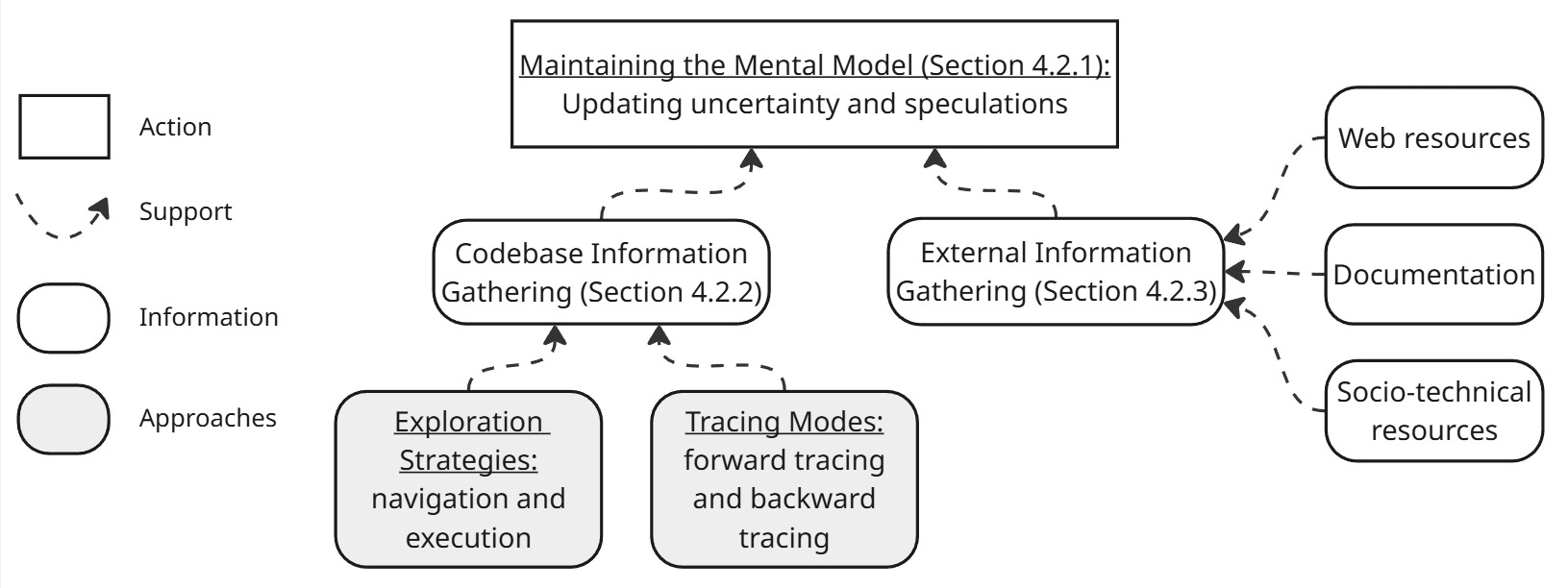}
    \caption{Mental Model Development Process. Developers maintain a mental model of uncertainty and speculations by gathering codebase and external information.}
    \label{fig:mental_model}
\end{figure}

Developers have long been known to create and maintain mental models of software systems~\cite{LaToza2006:Maintaining}. At the start of a debugging task, developers often hold a partial (reflecting limited familiarity with the codebase) and sometimes incorrect (explaining the presence of the bug) mental model. Debugging is therefore an iterative process in which developers assess their own uncertainty, identify tactics to reduce uncertainty, and extend or correct their mental models. Regarding this process, P5 explained, ``The most time I typically spend on fixing bugs is the finding step and understanding what is wrong.'' This account shows that debugging is as much about reasoning and reconstruction as it is about implementing a fix, a theme we observed across all developers. We identify two aspects of the mental model that a developer maintains: their uncertainty (what they believe they do not know) and their speculations (what they presume might be happening). To update the mental model, developers gather information from two sources: within the codebase and externally (\cref{fig:mental_model}). In Section ~\ref{sec:maintain_mental_model}, we explain at a high level how the mental model is maintained while minimizing effort. Then, in the following sections, we expand on \textbf{codebase information gathering} techniques through the lens of exploration strategies and tracing modes, and discuss \textbf{external information gathering} techniques of consulting a number of different sources.
 
\subsubsection{Maintaining the Mental Model}
\label{sec:maintain_mental_model}
This model builds on hypothesis-driven approaches from prior literature because this process emphasizes uncertainty and speculation, using vague statements that reflect possible causes but which are not directly testable. The goal is to learn the \emph{least} about the codebase to fix the issue, instead of becoming knowledgeable in extensive aspects.

\paragraph{Updating uncertainty}
All developers tracked their uncertainty and updated it across all tasks. For instance, after using browser DevTools to examine the data mapping in storage, S4 remarked, ``that looks right, the extensions look right (…) so we stored it correctly, but then when do we return back?'' S4 then added a console log to check the data at a specific state in a function call. S4 updated the mental model by tracing execution, reducing uncertainty about the system state and directing attention to the next point of inspection. Similarly, P7 used a debugger to step through functions, reading the comments, function names, and lines of code and making claims like ``this is the snippet that we need'' and ``this is where we handle [data].'' Updating uncertainty often relies on factual evidence (or lack thereof) rather than hypothetical reasoning: developers focus on clarifying what is known and unknown about the codebase, delimiting the boundaries of reliable knowledge, thereby creating a more stable foundation for speculative reasoning or making plans.

\paragraph{Updating speculations}

Participants navigated their uncertainty by generating and testing speculations. None of the developers articulated formal hypotheses; instead, they voiced casual thoughts about the system and evaluated whether those should guide the next actions. P5 explained:
\begin{quote}
    ``If I am working in the full stack, I start at the top of the stack and work my way down. So, I start with the HTML, go into the React components, go on the front end, then make my way to the APIs, and then I will work towards the back end.''
\end{quote}
This systematic traversal was also how P5 approached task B, despite having different speculation of ``component conflicts,'' a ``submit bug,'' a ``backend parser'' issue, and a ``connection issue.'' While these were all feasible, P5 did not immediately check these speculations, ``just to try to not to jump to conclusions.'' Here, P5 shows how speculation can function less as a direct driver of action and more as a background set of possibilities to consider, while structured traversal and examination of the system provided a systematic method of updating uncertainties. Using this approach, P5 eventually became confident that the problem stemmed from the TypeScript \textit{spread} operator. While navigating the codebase and checking different components, P5 generated speculations at multiple levels of abstraction. While most proved unrelated, these speculations illustrate active engagement with the bug and the exploration of alternative explanations. 

\paragraph{Breadth-first mental model development} 
In contrast with some prior advice on debugging, which suggests that developers should obtain thorough knowledge first~\cite{AgansDavidJ2006DT9I}, we observed that developers carefully \emph{avoided} tactics that would require that they learn too much, since in complex systems this learning process could be very costly. P1, for example, invested 52 minutes in finding out which change had introduced the bug, 12 minutes in looking for previous fixes, and 13 minutes in reproducing the issue at a different scope, since that might be more efficient than carefully investigating the cause of the erroneous behavior directly. Similarly, P7's choice of using AI to help understand and debug certain components of the code, S2 consulting experts, and P3 applying temporary DOM fixes before moving to implementing the changes in the codebase, can all be seen as efforts to avoid the never-ending effort of understanding the system. These strategies highlight how developers often aim to construct a ``good-enough'' mental model that allows them to move forward, rather than striving for exhaustive understanding. In doing so, they balance the need for accuracy with the practical goal of minimizing cognitive and temporal effort during debugging. 

\subsubsection{Codebase Information Gathering}
\label{sec:gather_information_codebase}

What information do developers seek inside the codebase? In our study, even developers familiar with the project often could not recall exact file locations, function names, or concrete implementations. Several developers with years of experience also reported limited familiarity with a specific framework, tool, or file structure. P7 explained, ``When I go to a function, I find something more.'' This illustrates debugging as an information-gathering process in which programmers progressively uncover or recall details in the code. Codebase information gathering involves reasoning at multiple levels of abstraction: locating files and entry points, interpreting interfaces and contracts, tracing data flow, and understanding configuration or environment factors. Prior experience helps shape systematic approaches: where to look first, how to form hypotheses, and which cues are most diagnostic. Many participants linked their ability to make immediate jumps in the codebase to this accumulated experience, often framing their reasoning with phrases like ``I know.'' In this way, codebase information serves to contextualize the directly accessible information that developers draw on while debugging.

To understand how developers gather codebase information, we consider two complementary dimensions. The first is \textbf{exploration strategies}, which describe how programmers engage with the system: by navigating its static structure (e.g., opening files, reading definitions) or by executing code to observe runtime behavior (e.g., stepping through, printing values). The second is \textbf{tracing modes}, which describe how programmers reason about cause and effect. Consistent with prior work on debugging strategies~\cite{Weber2025:Understanding}, we identified two modes: forward tracing (predicting outcomes from a cause) and backward tracing (reasoning back from an observed failure). These two dimensions are analytically distinct but often intertwined. In the rest of this section, we define these concepts precisely and illustrate them with examples.

\paragraph{Exploration Strategies}
\label{sec:exploration}
We distinguish two exploration strategies used during debugging. \textit{Navigation} is static inspection of the code without running it (e.g., reading code, project‐wide search, ``go to definition,'' symbol lookup). \textit{Execution} involves running code to observe behavior, and is often coupled with instrumentation and runtime manipulation. For example, P6 and P7 traced calls with breakpoints, P2 and P5 used console logs, P4 monitored calls with analysis tools, and P1 used \texttt{git bisect} with a test script. These interventions help gather concrete runtime information and are often associated with debugging strategies in literature about how software engineers approach debugging~\cite{arab2025developerschoosedebuggingstrategies,Liu2023}. In Section~\ref{sec:debugging_techniques}, we list the strategies we identified. 

Combining navigation with execution was nearly universal. There were only two exceptions: P4, who could not navigate due to restricted access to API calls and source files, and S1, whose syntax bug was resolved without executing the program. However, there are patterns in strategy application.
First, \emph{programmers often start by navigating to relevant files} for further information gathering after reproducing an issue. To navigate the files and find relevant code in a codebase with which they are only partly familiar, developers often use information they gather from reproducing the bug. P3 and P5 both started to examine the codebase by inspecting the webpage to identify a key word they could use to find relevant files. In task A, P6 inspected a global configuration object to trace the array of JavaScript, using this information to find the files. 

On the other hand, \emph{developers rely on execution more often in complex issues}. S5 repeatedly alternated between reading code and running it: quickly jumping to function definitions to understand where and when functions were invoked, articulating expectations about behavior, then setting breakpoints to inspect variables at runtime once a suspicious region was found. For instance, when reading the code without execution, S5 moved between several functions and made claims about how the program works, such as ``they are already loaded,'' and ``I am constructing this… I think it is in there that I load the icon…'' At one point during this inspection, S5 commented ``I guess we are saying that the icon is not set here? [Let me] set a breakpoint here…'' and switched to execution while monitoring the variables, concluding after observing a conditional breakpoint that ``the list has icons here.'' This switch dominated the entire 2-hour debugging process, where S5 moved to several different relevant functions, made speculations, and investigated whether a particular function was the root cause. Similarly, P6 mentioned that \emph{execution} offers the concrete evidence necessary as part of information gathering. When reading the code, P6 said, ``We are getting patient ID here from this function (…) something is wrong, but without a debugger, I cannot [be sure].'' Using a debugger, P6 identified an issue with the regular expression. Execution enabled P6 to gather reliable information instead of speculating and helped narrow down to a general region where P6 speculated the issue was. 

\emph{For simple bugs, execution beyond reproduction and testing may not be necessary.} Despite using console logs to monitor calls during tasks B and C, P5 did not execute the code during the debugging phase for task A, which P5 considered ``really simple.'' Instead, P5 immediately implemented the fix after reading the function. Despite debugging in the same codebase for tasks B and C, P5 treated task A as straightforward and relied less on execution, reinforcing the broader pattern that programmers \emph{turn to execution more often when dealing with complex issues}, enabling checking on speculations.

\paragraph{Tracing Modes}
\label{sec:tracing}
Navigation and execution are guided by a choice of \emph{forward tracing} or \emph{backward tracing}. Forward tracing reasons through code by following the logical flow of code execution from an initial known state or input. Seven of the eight participants in our study used this strategy, particularly during fine-grained code reading, where developers examined small code segments to infer behaviors and locate issues. For example, P5 ran the code while observing console logs for output. P4 debugged a performance issue that involved extreme page loading delays and browser crashes. Unlike defect reports with concrete expected and actual behaviors, this issue lacked a clear failure signature. P4 used Chrome DevTools~\cite{chromedevtools_docs} to monitor the page load process, collecting performance traces and observing the sequence of API calls. Here, forward tracing enabled gathering empirical information to form a diagnosis of the issue based on the observed behavior.

In contrast, \textit{backward tracing} begins with the issue and works backward to identify the conditions or changes that could have caused it. Backward tracing emerged as a dominant strategy, particularly at coarser-grained levels such as version history, files, and function calls. It is often used to rapidly reduce the search space in a large codebase. For example, P1 used \texttt{git bisect} combined with testing scripts. Although executing test scripts follows the forward execution of a program, it represents backward tracing in P1's context because the action is used to support a time-wise backward tracing effort to localize the faulty commit.

These two modes correspond with Spinellis's \emph{Drill Up from the Problem to the Bug or Down from the Program’s Start to the Bug}~\cite{Spinellis2016:Effective} methods. According to Spinellis, backward tracing should be used when the issue has an identifiable cause, and forward tracing should be used for failures that are difficult to pin down. However, we see the tracing modes as a dynamic process that also depends on the abstraction level and the codebase complexity. Many participants began root cause analysis with backward tracing (P2, P3, P5, P6), typically navigating the codebase at a higher file or module level. As they narrowed down to the exact function that might be causing the issue, they shifted to a more granular view, reading the implementation line by line to understand the specific behavior. For example, P3 quickly located a relevant file via backward tracing, saying, ``[reads variable assigned value] that should probably be in [file name].'' Using this inference, P3 searched for and opened the file containing the issue. However, moments after entering it, P3 switched to forward tracing, reading the function implementation to determine whether it required a fix: ``There is some other things in here (…) [reads through function implementation] but I do not think we are using it (…).'' This shift from backward to forward tracing often occurs when the problem scope is roughly known, and the programmer has a mental model of the system’s structure but lacks sufficient implementation detail. When asked about familiarity with the codebase, P3 remarked, ``I am not familiar with this codebase (…) [but] we have similar codebases and because of the naming convention, I could quickly go to the file.'' Leveraging the naming convention and backward tracing allowed P3 to locate a relevant area, but reading through the code line by line was still necessary.

Developers' tracing mode depends on codebase complexity and familiarity. Forward tracing was dominant in high-familiarity codebases (P2, P5), while backward tracing was more common in low-familiarity and longer tasks (P1, P3, P4). P1, who reported less than 5\% familiarity with a very large codebase, was the only participant who relied on backward tracing as the primary mode of evidence-gathering. P3 and P4 (both with $<20\%$ codebase familiarity) used backward tracing as a secondary strategy, and P2 and P5 (both with $\geq50\%$ codebase familiarity) used backward tracing less often.

Participants in low-familiarity contexts tended to avoid forward tracing because it requires in-depth reading of small code segments. Forward tracing often involves examining code within a single file or function, but neither P1 nor P4 opened any files to inspect full code implementations. P1 focused on version history and commit changes, relying on these fragments to infer program behavior without directly reading the code. Likewise, P4 concentrated on structural analysis. Although remarking ``this piece of code is executed multiple times (…),'' P4 never opened the relevant files. Instead, P4 focused on understanding the broader system-level effects of that code.

In contrast, P2 and P5 were already familiar with the codebase and showed no concern in directly diving into the code. Compared with the personal project used to demonstrate debugging techniques, P5 noted how a company debugging task might involve more searching: ``Sometimes I will actually say the most time I typically spend on fixing bugs is the finding step (…) in this case, I kind of know what's wrong because I built it. It is a lot harder when you are not the one who actually touched the code. It is a lot more of an archaeology.'' From the collective experience of the developers, this archaeology might involve more frequent backward than forward tracing.

\subsubsection{External Information-Gathering}
\label{sec:gather_information_external}

While much of programmers' debugging work involves interrogating the codebase, participants also turned to external sources when direct inspection was insufficient or inefficient. These resources complemented codebase evidence by providing technical or social context and offered access to knowledge beyond the immediate codebase environment. 
Developers updated their mental models by consulting web resources, making human connections, and reading relevant documentation or related repositories.

\paragraph{Web resources}

Web resources enabled programmers to draw on external expertise when local inspection alone was insufficient or time-consuming. In debugging task B, a packaging issue that P2's mentee had struggled with, P2 quickly recognized the error after reproducing it (``I have seen this error before''), turned to web searches, and quickly found a working resolution through Stack Overflow. Leveraging awareness of likely failure reasons and precision in describing the problem, P2 refined the search query from ``resolve differences in version'' to ``compatibility.'' By relying on web resources to locate the exact solution, P2 demonstrated using the broad web knowledge base while minimizing local debugging. Searches bridged between the immediate problem space and wider communities of practice, expanding the scope of available debugging evidence.

Participants treated AI-driven tools as an extension of web-based information-gathering. S1 used Claude Sonnet~\cite{claude-sonnet} to inquire about the struct that caused the syntax error. Referencing the sample code, S1 was able to fix the syntax error more efficiently than after longer unsuccessful attempts to consult the documentation. P7 used IDE-embedded AI extensively for debugging, highlighting that exploring the use of LLMs was encouraged in their company. For example, to explain a specific segment of code, P7 highlighted the segment and prompted ``what exactly is this piece of code doing to formatting?'' P7 described the benefits using LLMs for debugging:
\begin{quote}
    ``When it comes to stuff like [rich text syntax and parsing] (…) we can use AI to make sure we do not have to waste too much time going through documentation for things that are very, very small, like this one.''
\end{quote}
Despite mentioning LLMs' power in working across files and web resources and providing expertise in unfamiliar fields, P7's attempts at using AI to generate a fix were not effective, eventually fixing the issue manually. P7 observed, ``The thing is that since it is AI-generated, we do not actually know how to handle [a rich text format],'' suggesting that AI might leave code that is \textit{harder} to maintain and debug.

Developers integrated web-based resources selectively. As the examples above show, both LLMs and traditional search tools support debugging especially when developers had relevant expertise but lacked specific technical details. In such cases, web searches or AI queries fill the gap by surfacing fragments of knowledge without extensive documentation review.

\paragraph{Socio-technical resources}

Programmers drew on interactions with other people and collaborative artifacts that embed social knowledge --- as part of their debugging process. For example, some streamers engaged directly with chat participants to brainstorm solutions with viewers in real time (S3, S5). S3 identified the root cause during this interaction, illustrating the potential of pair programming-like collaboration in debugging. Meanwhile, S2 reached out to another maintainer on Discord to ask targeted questions about what issue was triggered in which portion of the codebase. P1 similarly identified colleagues with relevant expertise and redirected the issue to those best equipped to resolve it. Despite showing a personal project, P5 observed, ``being able to go to commits and see what people actually did and how that happened gives me a lot of context.''

Some developers also faced challenges for socio-technical reasons in which they could not access the relevant code, data, or environment needed for debugging. These restrictions limited their ability to perform an in-depth investigation or reproduce issues, or even becoming the root cause of the issue. This was especially evident among participants who work with code they had not originally written (P1, P3, P4, P6, P7, S2), including in company projects with authentication concerns. For example, P4 had to analyze the performance issue without access to the code, mentioning,
\begin{quote}
``One way is to go [to the] platform team, and talk to them about how their APIs are performing bad, [and discuss] what we can do about them. (…) looking at this, we do not know what it is we are requesting and what is it we are getting. ''
\end{quote}
Because of these challenges, P4 had to ``go by intuition'' and do ``trial and error,'' which they expressed were not the optimal debugging conditions, and expressed how they hoped to see changes in the development workflow. Similarly, both P3 and P6 were blocked by authentication issues. Because of this, P6 had to switch to a different task to show, while P3 had to schedule a follow-up session after asking for credentials. P1 spent considerable time investigating a bug that turned out already fixed, but debugging was challenging because another contributor had used the version control system in an unusual way, resulting in reapplying an old patch with the same commit message.



\paragraph{Consulting documentation and related codebases}
Participants frequently consulted documentation or related codebases to fill gaps in their understanding. This included referring to syntax documentation within the IDE (S1), checking guides for external service integration and comparing them with corresponding code (S2), and examining documentation to clarify the structure of parsed items (S4). Use of documentation appeared more common among streamers than traditional think-aloud participants. One possible explanation is that streamers' projects often embodied stretch learning goals or intentionally challenging domains, which may have made documentation a more valuable and accessible support than relying solely on experimentation or peers. Another contributing factor could be the performative nature of streaming itself: consulting documentation provides a visible way of engaging with uncertainty, which aligns with the pedagogical or entertainment aspects of streaming.

\subsection{Factors Influencing the Debugging Process}

\label{sec:influences}

Developers' experience and awareness of version history both influence the debugging process.

\subsubsection{Experience}
Professional programmers drew heavily on prior experience and familiarity with the codebase to navigate debugging tasks efficiently. In our study, all participants showed how experience informed their hypotheses. In some cases, they leveraged this experience to bypass information-gathering steps and move directly to conclusions.

Prior knowledge and experience helps developers form hypotheses and, at times, take shortcuts to a solution. Guided by past experience with a similar packaging issue, P2 quickly resolved a reported bug that junior developers on the team had struggled with:  
\begin{quote}
    ``[Looks at error message in the terminal] I think someone with a different node value has probably pulled the project and committed… I am very sure that there will be a discrepancy (over) the versions of the libraries we are using there, that is why they are getting this error… I have experienced this before.''
\end{quote}
This prior experience allowed P2 to immediately propose a root cause rather than tracing or inspecting additional error messages. P2 then identified the solution through web search. Experience enabled P2 to reduce the debugging process to a single web search.

Prior knowledge can be applied more generally. P3 showed how patterns, conventions, and structural similarities can be helpful. When asked about familiarity with the specific codebase, P3 explained, ``…actually, no. I am not familiar with this codebase… we have similar codebases, and because of the naming convention, I could quickly go to the file.''   

In addition to personal experience, some developers leverage the experience of others. P5 described this practice: ``When I come into a company, I will ask if someone will show me how they found their bug … I pick out pieces from them of how I can approach it.'' P1 also demonstrated the use of vicarious experience. Rather than immediately inspecting the code, P1 spent time reviewing reports of similar issues to learn how they had been handled. After about ten minutes, P1 concluded, ``I don't see anybody has fixed this.'' Although these reports did not resolve the bug, this method enabled P1 to rule out simpler solutions before investing effort in more complex approaches.

\subsubsection{Version Awareness}
In large codebases, version history provides insight into how and why the system has evolved, supporting reasoning about possible causes of an error. The impact of versioning was more evident in a large, collaborative codebase compared to small, personal projects. P1's major concern was finding ``(the) commit that either introduces the issue, or that combined with my changes introduces the issue.'' Both the \emph{timing} and the \emph{method} mattered in the process of debugging. P1 studied commit messages, comparing versions through git diff, and using git bisect to find the commit in question.

\section{Use of Debugging Techniques}


\label{sec:debugging_techniques}

\begin{table}
  \caption{Debugging Techniques Codebook and the Count Across Unique Tasks (Out of 17 Tasks). }
  \small
  \label{tab:debugging_techniques}
  \begin{tabularx}{\linewidth}{p{2.5cm}|p{7.2cm}|>{\RaggedRight\arraybackslash}X}
    \toprule
    \textbf{Name} & \textbf{Definition} & \textbf{Application in Task} \\ \hline
    \midrule
    line-by-line reading & Read lines of code, typically within a single function. & P1TA, P2TA/B, P3TA, P5TA/B/C, P6TA/B, P7TA, S1TA, S2TA, S3TA, S4TA/B, S5TA\\ \hline
    navigate files & Search and look across the file structure. & P2TA/B, P3TA, P5TA/B/C, P6TA/B, P7TA \\ \hline
    navigate functions & Search and trace across the functions, sometimes between files. & P2TA, P3TA, P5TB/C, P6TA/B, P7TA, S4TA/B, S5TA\\ \hline
    \makecell[tl]{navigational\\historic analysis} & Exercise version awareness using tools such as git diff to statically understand the previous versions of code. & P1TA, P6TB \\ \hline
    executional historic analysis & Exercise version awareness using tools like git bisect that analyzes historical versions by running the code. & P1TA \\ \hline
    \makecell[tl]{run tests or\\benchmarks} & Run new or pre-defined tests to observe pass, failures, or performance. & P1TA, P3TA, P7TA, S3TA\\ \hline
    read error message & Read error messages in the terminal, console or on the webpage to understand the issue. & P1TA, P2TB, P3TA, S1TA, S2TA, S4TA/B\\ \hline
    \makecell[tl]{trigger DOM/UI\\changes} & Perform actions like clicking a button to trigger the issue. & P3TA, P4TA, P5TA/B/C, P6TA/B, S2TA, S4TA/B\\ \hline
    DOM inspection & Inspect DOM-related settings, typically using DevTools. & P3TA, P5TA/B/C, P6TA/B, S4TA/B \\ \hline
    \makecell[tl]{setting\\configuration} & Configure developer settings like timeout threshold or turn on debugger mode for better debugging experience. & P4TA, P6TB, S2TA, P1TA\\ \hline
    API analysis & Analyze API calls through inspecting DevTools or using API platforms. & P4TA, P5TB, P6TA\\ \hline
    \makecell[tl]{printf-style\\debugging} & Use statements like print, printf, or console log to quickly monitor processes or variables without interrupting the execution. & P2TA, P5TB/C, P6TA, P7TA, S2TA, S4TA \\ \hline
    debugger & Use debugger functionalities to extensively examine and explore the code execution, variables, and data. &  P6TA/B, P7TA, S4TA, S5TA \\\hline
    documentation & Check API or syntax documentation related to the bug. & S1TA, S2TA, S4TB\\ \hline
    social-technical sources & Seek help from knowledgeable parties. & S2TA, S3TA\\ \hline
    web sources & Seek help from web resources and AI. & P2TB, P7TA, S1TA\\ \hline
    \bottomrule
  \end{tabularx}
\end{table}

~\Cref{tab:debugging_techniques} summarizes the debugging techniques. Backed by the grounded theory framework, we categorize debugging techniques based on \textit{actions} and associate actions with the \textit{intentions} of using the techniques in the technique definitions. We also identify that the use of debugging techniques can be complex, multi-layered actions. For example, line-by-line code reading can be embedded in the use of tools such as git diff or the debugger.



The most prevalent debugging technique we observed was \textit{line-by-line reading} of code (16), followed by \textit{navigate functions} and \textit{trigger DOM/UI changes} (10). \textit{Navigate files} (8), \textit{DOM/UI inspection} (7) were also common. Navigating the code at the line, function, and file levels is often aided by execution strategies such as \textit{trigger DOM/UI changes}, \textit{printf-style debugging}, as well as the use of \textit{debugger}. 

Each task used an average of 5.5 techniques, and each developer used an average of 5.9 techniques. A correlation analysis revealed a weak positive association between time and count (Spearman's $\rho=0.10$), which was not statistically significant ($p=0.68$).

\section{Discussion}
From our theory, we extract design opportunities and pedagogical implications.

\subsection{Design Opportunities}

Our findings suggest several directions for designing future debugging and software engineering tools. These directions reflect how programmers manage uncertainty, integrate socio-technical knowledge, and debug with AI-based assistance.

\paragraph{Supporting mental model management}
Developers in our study treated mental model management as central to debugging. They continuously revised their knowledge of the system. However, tools provide limited support for tracking these shifting states of knowledge and strategies used to extend them~\cite {latoza2020explicit}. Future tools could make uncertainty management more explicit by tracking untested speculations. For example, IDE tools can allow for tracking the history of files and functions that the developer looked at to record the exploration trail. Future debugging tools can also represent levels of confidence across the code or identify which regions of the program can be ruled out. In line with the knowledge-avoidance strategies we observed, tools should help developers make progress without requiring them to internalize large portions of the system.

\paragraph{Supporting socio-technical integration}
Debugging in complex codebases and organizational settings was rarely an individual activity. Developers drew on version histories, prior fixes, colleagues' expertise, external communities, and documentation. In several cases, socio-technical barriers such as limited access to code or unusual version control practices slowed or prevented progress. Addressing these issues will require both tool design (e.g., systems that make expertise and history more accessible) and organizational effort (e.g., norms and standards for documentation, version control, and data sharing). Our study highlights the importance of designing debugging support that integrates social and technical resources.

\paragraph{AI-informed Debugging Practice}
The selective use of AI assistants points to opportunities for improving AI-informed debugging. In our study, developers used general-purpose LLMs rather than domain-specific tools. To better situate LLMs for debugging purposes, AI systems for debugging could provide transparent rationales for their suggestions, expose uncertainty in their recommendations, cite authoritative sources, and support hand-offs between automation and human reasoning.  

\subsection{Pedagogical Implications}
Our findings contribute to a computing theory that research in computing education can draw on, and highlight challenges and opportunities in teaching debugging~\cite{Chen2019:Towards,McCauley01062008, computing_ed_today_2019}.  Debugging is difficult to master and teach. Many factors that developers rely on, such as prior experience and familiarity with the codebase, cannot be directly transferred to classroom settings. However, our observations highlight practices that can be intentionally scaffolded: guiding the iterative updating of mental models and preparing novices to debug complex scenarios. Researchers can also use our theory to analyze student debugging practices and identify potential gaps in student practices compared to professionals.

\paragraph{Encouraging reasoning through uncertainty and speculation} Debugging pedagogy often emphasizes systematic hypothesis-driven strategies~\cite{zeller2009programs}. While valuable, our study shows that professional debugging emphasizes reasoning through uncertainty and speculation rather than testing well-formed hypotheses. Developers updated their mental models by identifying what they know and do not know, and by carefully evaluating tentative speculations. Educators should therefore teach students to manage uncertainty, avoid premature commitment to a single explanation, and evaluate speculative reasoning systematically. Early exercises can scaffold students in articulating what they know and do not know about the system, plan information-gathering steps at different levels of abstraction, and delay commitment to implementing a fix until sufficient evidence is gathered. Assignments can focus on scaffolding students in using techniques like backward tracing through actionable debugging steps to formalize the process of learning debugging.

\paragraph{Training in complex, unfamiliar codebases}
Although the high-level debugging process was consistent across settings, codebase information gathering differed between simple and complex systems. Students often work on small, self-contained assignments, which limits opportunities to practice version awareness, consult socio-technical resources, or navigate unknown technical areas. Curricula should therefore incorporate debugging tasks in real, complex codebases, normalizing uncertainty management and the coordination of multiple strategies. This approach would better prepare novices for professional practice, where debugging often involves both technical and organizational challenges.

\section{Related Work}
Debugging has long been recognized as a central component of software development~\cite{Vessey}, comprising a significant part of development costs. In this section, we summarize prior work that has identified a variety of specific debugging strategies that developers use.

\paragraph{Studies of Debugging}
A literature review of 74 papers~\cite{Weber2025:Understanding} identified 12 different strategies. Early work with students on small codebases identified techniques such as \textit{forward reasoning}, \textit{backward reasoning}, and \emph{following execution}, which extends \emph{forward reasoning} by using a debugger to visualize state~\cite{Katz1987:Debugging,ROMERO2007992}. Later, Bohme et al. asked 12 software engineers to fix 27 bugs extracted from a corpus, finding that they used a combination of \textit{forward reasoning}, \textit{backward reasoning}, \textit{code comprehension} (reading the code), \textit{input manipulation} (constructing a similar test case to compare the behavior), \textit{log analysis}, and \textit{experience}~\cite{Bohme2017:Where}. Among these, \textit{forward reasoning} and \textit{code comprehension} were used most frequently.

In interviews, developers reported that \emph{hypothesis generation and testing} formed a baseline method~\cite{arab2025developerschoosedebuggingstrategies,Layman2023:Debugging}. These hypothesis generation approaches relate to \textit{scientific debugging}~\cite{zeller2009programs}, which involves generating, testing, and revising hypotheses about causes of bugs. This perspective is consistent with results from interviews of eight professional developers, who reported using variants of the scientific method.~\cite{Siegmund2014:Studying}. A follow-up survey found that in the most difficult bugs to fix, the distance between the root cause and the incorrect behavior was especially large~\cite{perscheid2017studying}. Participants reported spending 20\% -- 40\% of their time debugging. 

In a study of 12 participants who debugged the same performance bug in a database system while using a set of VR-based developer tools, the most frequent strategy was \emph{source code inspection}, followed by \emph{following data and control flow}~\cite{Weber2025:Understanding}. A \emph{scientific approach}, involving stating and testing a hypothesis, was used only rarely (79 15-second instances, compared with 832 instances of inspecting source code). Not surprisingly, the distribution of strategies for debugging performance problems may differ from those used in general practice. However, because all participants in the study worked on the same bug, it is difficult to generalize the results. 


Developers have also reported using a \emph{replication-observation-deduction} pattern~\cite{Hirsch2021:What}, consistent with our findings, as well as backtracking (likely corresponding with backward reasoning) and code inspection. In the context of more general programming tasks, professionals often avoid program comprehension, instead applying structured comprehension strategies based on experience, codebase information, and external information~\cite{RoehmSoftwareComprehension}. In that study, only 10 of 28 tasks were debugging tasks, suggesting that our observations of \emph{avoidance} generalize beyond debugging to other programming tasks.

More general theories of information-finding may also apply to debugging. \emph{Information foraging theory} models the search for information as a predator searching for prey by following scent~\cite{Lawrance2013:Programmers}. When applied to debugging, this theory predicts that developers use information to guide a search rather than suggesting a hypothesis-driven, scientific approach~\cite{Brooks1983:Towards}. 

Other work has shed light on the \emph{actions} that developers took in debugging portions of live-streamed programming tasks, finding that debugging involves 41\% editing and 29\% testing~\cite{debugging_episodes}. Comparing debugging and implementation work, debugging requires more time inspecting, but implementation requires more time studying external resources. Debugging may also require more frequent context switching and higher cognitive load than implementation.

Finally, although our work focuses on general-purpose debugging strategies, specific contexts, such as frameworks, require special debugging techniques. Coker et al. found that debugging software that uses frameworks depends on the ability to manage frameworks' inversion of control, complex object protocols, and the complex relationships between static and dynamic artifacts~\cite{CokerFrameworkDebugging}.

\paragraph{Analysis-Based Tools}
Based on program analysis approaches, researchers have proposed \emph{automated debugging} techniques, many of which recommend suspicious lines of code to review via program analysis or statistical methods. Nonetheless, developers still fix most bugs manually; one study of automated debugging tools found a modest benefit for novices but not a general improvement~\cite{Parnin}. Studies of IDE-based debugging tools found that only a minority of developers still use these tools, with the rest relying on traditional means such as print statements~\cite{Beller2018:Dichotomy}.

\paragraph{Expert advice}
Textbooks provide expert guidance regarding debugging methods~\cite{AgansDavidJ2006DT9I,Spinellis2016:Effective} and describe how to use specific tools~\cite{stallman1988debugging}. Although the grounding of these recommendations in evidence is often unclear, these recommendations reflect common advice among practitioners. Agans, for example, argues that one should first understand the system in great depth, e.g., by exhaustively reading manuals~\cite{AgansDavidJ2006DT9I}. In contrast, we found that the complexity of software systems often required \emph{avoiding} reading because it is impossible to read everything. Spinellis gives 66 techniques that might be used in updating mental models~\cite{Spinellis2016:Effective}. Some of the high-level strategies, such as backward reasoning, correspond with those described in earlier empirical work; others reflect domain-specific techniques that are helpful in particular scenarios for implementing the high-level strategies.

\paragraph{Our study}
Our grounded theory approach contrasts with prior work in that it provides a holistic perspective, identifying both high-level approaches (iterating around a mental model) and specific strategies. There appears to be a contrast between the scientific strategies that many developers report using in interviews and personal accounts~\cite{Siegmund2014:Studying,Eisenstadt1997:Hairiest} and observational studies and experiments, whose findings support use of other strategies more strongly~\cite{Weber2025:Understanding,Katz1987:Debugging,ROMERO2007992,Bohme2017:Where,Whalley2011:Novice,RoehmSoftwareComprehension}. This discrepancy may be explained by our findings, which show that developers often express uncertainty and make speculative statements, which are not quite hypotheses but express beliefs and informal knowledge that resemble hypotheses. Hypotheses are known to be valuable; correct hypotheses predict debugging success~\cite{Alaboudi2020:Hypotheses}.

In contrast with previous large studies, our work observes professionals working on a variety of codebases that were familiar to them, providing insight that may be more applicable to professional practice and more representative of the challenges of everyday software engineering work. Our approach enabled us to ask the participants questions about their strategic choices, affording us insight that we may not have obtained via observations alone. This study significantly extends our preliminary work~\cite{Liu2023}, which had a limited sample size and did not develop a theory.

\section{Future Work and Conclusion}
We present a grounded theory of professional debugging as an iterative process centered on mental model development. Developers reduce uncertainty and generate speculations by combining exploration, tracing, and external resources, often aiming for a ``good-enough'' understanding rather than exhaustive knowledge. We suggest design opportunities for tools that track uncertainty and integrate human knowledge, and highlight teaching approaches that prepare students to update mental models in complex, unfamiliar systems. Building on our theory, future work could conduct longitudinal studies on how debugging relates to the broader software engineering workflow, compare debugging process patterns across different domains, or evaluate how tools or AI affect mental model development.

\section{Data Availability}
We provide a partial replication package for this study. The package includes publicly accessible links to the streamer video recordings and corresponding codebases, which were accessible at the time of analysis, along with the video segments identified as debugging moments. We also provide the closed-coding results for the debugging processes discussed in Sections~\ref{sec:debugging_techniques} and~\ref{sec:theory_overview}. Open coding data from the grounded theory analysis is not publicly available, as they contain sensitive information that could compromise participant privacy. However, an example illustrating the coding process is provided in Section~\ref{sec:methods}. Finally, the study protocol script is included as a guide to the semi-structured observation and interview process. These materials are available at:
\url{https://github.com/HaileyHL/grounded_theory_debugging_supplementary_material}.

%
\begin{acks}
The authors appreciate the help of the companies and participants for making this study possible. In addition, we thank the work of the anonymous reviewers for providing helpful suggestions on improving the paper.
\end{acks}

\bibliographystyle{ACM-Reference-Format}
\bibliography{citations}

@book{Charmaz2014:Constructing,
  author    = {Charmaz, Kathy},
  year      = {2014},
  title     = {Constructing Grounded Theory},
  publisher = {Sage},
  edition   = {2nd.}
}

@misc{arab2025developerschoosedebuggingstrategies,
      title={How Developers Choose Debugging Strategies for Challenging Web Application Defects}, 
      author={Maryam Arab and Jenny T. Liang and Valentina Hong and Thomas D. LaToza},
      year={2025},
      eprint={2501.11792},
      archivePrefix={arXiv},
      primaryClass={cs.HC}
}

@article{grounded_copilot,
author = {Barke, Shraddha and James, Michael B. and Polikarpova, Nadia},
title = {Grounded Copilot: How Programmers Interact with Code-Generating Models},
year = {2023},
issue_date = {April 2023},
publisher = {Association for Computing Machinery},
address = {New York, NY, USA},
volume = {7},
number = {OOPSLA1},
url = {https://doi.org/10.1145/3586030},
doi = {10.1145/3586030},
journal = {Proc. ACM Program. Lang.},
month = apr,
articleno = {78},
numpages = {27}
}

@article{debugging_episodes,
author = {Alaboudi, Abdulaziz and LaToza, Thomas D.},
title = {What constitutes debugging? An exploratory study of debugging episodes},
year = {2023},
issue_date = {Sep 2023},
publisher = {Kluwer Academic Publishers},
address = {USA},
volume = {28},
number = {5},
issn = {1382-3256},
url = {https://doi.org/10.1007/s10664-023-10352-5},
doi = {10.1007/s10664-023-10352-5},
journal = {Empirical Softw. Engg.},
month = sep,
numpages = {34}
}

@article{Liu2023,
author = "Amanda Liu and Michael Coblenz",
title = "{Debugging Techniques in Professional Programming}",
year = "2023",
month = "3",
url = "https://kilthub.cmu.edu/articles/conference_contribution/Debugging_Techniques_in_Professional_Programming/22277365",
doi = "10.1184/R1/22277365.v1"
}

@incollection{NASA-TLX,
title = {Development of NASA-TLX (Task Load Index): Results of Empirical and Theoretical Research},
editor = {Peter A. Hancock and Najmedin Meshkati},
series = {Advances in Psychology},
publisher = {North-Holland},
volume = {52},
pages = {139--183},
year = {1988},
booktitle = {Human Mental Workload},
issn = {0166-4115},
doi = {https://doi.org/10.1016/S0166-4115(08)62386-9},
url = {https://www.sciencedirect.com/science/article/pii/S0166411508623869},
author = {Sandra G. Hart and Lowell E. Staveland}
}

@book{AgansDavidJ2006DT9I,
author = {Agans, David J},
address = {Nashville},
edition = {1st.},
isbn = {9780814474570},
language = {eng},
publisher = {AMACOM},
title = {Debugging: The 9 Indispensable Rules for Finding Even the Most Elusive Software and Hardware Problems},
year = {2006},
}

@book{Spinellis2016:Effective,
author = {Spinellis, Diomidis},
title = {Effective Debugging: 66 Specific Ways to Debug Software and Systems},
edition = {1st.},
isbn = {9780134394909},
language = {eng},
publisher = {Addison-Wesley Professional},
series = {Effective software development series},
year = {2016},
}

@inproceedings{LaToza2006:Maintaining,
author = {LaToza, Thomas D. and Venolia, Gina and DeLine, Robert},
title = {Maintaining mental models: a study of developer work habits},
year = {2006},
isbn = {1595933751},
publisher = {Association for Computing Machinery},
address = {New York, NY, USA},
url = {https://doi.org/10.1145/1134285.1134355},
doi = {10.1145/1134285.1134355},
booktitle = {Proceedings of the 28th International Conference on Software Engineering},
pages = {492--501},
numpages = {10},
location = {Shanghai, China},
series = {ICSE '06}
}

@inproceedings{Bohme2017:Where,
author = {B\"{o}hme, Marcel and Soremekun, Ezekiel O. and Chattopadhyay, Sudipta and Ugherughe, Emamurho and Zeller, Andreas},
title = {Where is the bug and how is it fixed? an experiment with practitioners},
year = {2017},
isbn = {9781450351058},
publisher = {Association for Computing Machinery},
address = {New York, NY, USA},
url = {https://doi.org/10.1145/3106237.3106255},
doi = {10.1145/3106237.3106255},
booktitle = {Proceedings of the 2017 11th Joint Meeting on Foundations of Software Engineering},
pages = {117--128},
numpages = {12},
location = {Paderborn, Germany},
series = {ESEC/FSE 2017}
}

@article{Katz1987:Debugging,
author = {Irvin R. Katz and John R. Anderson},
title = {Debugging: An Analysis of Bug-Location Strategies},
journal = {Human–Computer Interaction},
volume = {3},
number = {4},
pages = {351--399},
year = {1987},
publisher = {Taylor \& Francis},
doi = {10.1207/s15327051hci0304\_2},
}

@article{ROMERO2007992,
title = {Debugging strategies and tactics in a multi-representation software environment},
journal = {International Journal of Human-Computer Studies},
volume = {65},
number = {12},
pages = {992--1009},
year = {2007},
issn = {1071-5819},
doi = {https://doi.org/10.1016/j.ijhcs.2007.07.005},
url = {https://www.sciencedirect.com/science/article/pii/S1071581907001000},
author = {Pablo Romero and Benedict {du Boulay} and Richard Cox and Rudi Lutz and Sallyann Bryant}
}

@book{zeller2009programs,
author = {Zeller, Andreas},
title = {Why Programs Fail: A Guide to Systematic Debugging},
year = {2009},
publisher = {Morgan Kaufmann},
edition = {2nd.},
address = {Boston},
isbn = {978-0-12-374515-6},
publisher = {Morgan Kaufmann Publishers Inc.},
doi = {https://doi.org/10.1016/B978-0-12-374515-6.X0000-7},
address = {San Francisco, CA, USA}
}

@INPROCEEDINGS{Siegmund2014:Studying,
  author={Siegmund, Benjamin and Perscheid, Michael and Taeumel, Marcel and Hirschfeld, Robert},
  booktitle={2014 IEEE International Symposium on Software Reliability Engineering Workshops}, 
  title={Studying the Advancement in Debugging Practice of Professional Software Developers}, 
  year={2014},
  volume={},
  number={},
  pages={269--274},
  doi={10.1109/ISSREW.2014.36}
}

@INPROCEEDINGS{Alaboudi2019:Exploratory,
  author={Alaboudi, Abdulaziz and LaToza, Thomas D.},
  booktitle={2019 IEEE Symposium on Visual Languages and Human-Centric Computing (VL/HCC)}, 
  title={An Exploratory Study of Live-Streamed Programming}, 
  year={2019},
  volume={},
  number={},
  pages={5--13},
  doi={10.1109/VLHCC.2019.8818832}
}

@article{perscheid2017studying,
  title={Studying the advancement in debugging practice of professional software developers},
  author={Perscheid, Michael and Siegmund, Benjamin and Taeumel, Marcel and Hirschfeld, Robert},
  journal={Software Quality Journal},
  volume={25},
  number={1},
  pages={83--110},
  year={2017},
  publisher={Springer},
  doi={https://doi.org/10.1007/s11219-015-9294-2}
}

@inproceedings{Parnin,
author = {Parnin, Chris and Orso, Alessandro},
title = {Are automated debugging techniques actually helping programmers?},
year = {2011},
isbn = {9781450305624},
publisher = {Association for Computing Machinery},
address = {New York, NY, USA},
url = {https://doi.org/10.1145/2001420.2001445},
doi = {10.1145/2001420.2001445},
booktitle = {Proceedings of the 2011 International Symposium on Software Testing and Analysis},
pages = {199--209},
numpages = {11},
location = {Toronto, Ontario, Canada},
series = {ISSTA '11}
}

@INPROCEEDINGS{Hirsch2021:What,
  author={Hirsch, Thomas and Hofer, Birgit},
  booktitle={2021 IEEE/ACM 8th International Workshop on Software Engineering Research and Industrial Practice (SER\&IP)}, 
  title={What we can learn from how programmers debug their code}, 
  year={2021},
  volume={},
  number={},
  pages={37--40},
  doi={10.1109/SER-IP52554.2021.00014}}

@article{Weber2025:Understanding,
author = {Weber, Max and Mailach, Alina and Apel, Sven and Siegmund, Janet and Dachselt, Raimund and Siegmund, Norbert},
title = {Understanding Debugging as Episodes: A Case Study on Performance Bugs in Configurable Software Systems},
year = {2025},
issue_date = {July 2025},
publisher = {Association for Computing Machinery},
address = {New York, NY, USA},
volume = {2},
number = {FSE},
url = {https://doi.org/10.1145/3717523},
doi = {10.1145/3717523},
journal = {Proc. ACM Softw. Eng.},
month = jun,
articleno = {FSE064},
numpages = {23}
}

@inproceedings{Whalley2011:Novice,
author = {Whalley, Jacqueline and Settle, Amber and Luxton-Reilly, Andrew},
title = {Novice Reflections on Debugging},
year = {2021},
isbn = {9781450380621},
publisher = {Association for Computing Machinery},
address = {New York, NY, USA},
url = {https://doi.org/10.1145/3408877.3432374},
doi = {10.1145/3408877.3432374},
booktitle = {Proceedings of the 52nd ACM Technical Symposium on Computer Science Education},
pages = {73--79},
numpages = {7},
location = {Virtual Event, USA},
series = {SIGCSE '21}
}

@article{lewis1982using,
  title={Using the "thinking Aloud" Method in Cognitive Interface Design},
  author={Lewis, Clayton},
  journal={Research Report RC9265, IBM TJ Watson Research Center},
  year={1982}
}

@inproceedings{stol2016grounded,
author = {Stol, Klaas-Jan and Ralph, Paul and Fitzgerald, Brian},
title={Grounded Theory in Software Engineering Research: A Critical Review and Guidelines}, 
year = {2016},
isbn = {9781450339001},
publisher = {Association for Computing Machinery},
address = {New York, NY, USA},
url = {https://doi.org/10.1145/2884781.2884833},
doi = {10.1145/2884781.2884833},
booktitle = {Proceedings of the 38th International Conference on Software Engineering},
pages = {120--131},
numpages = {12},
location = {Austin, Texas},
series = {ICSE '16}
}

@misc{chromedevtools_docs,
  title        = {Google Chrome DevTools Documentation},
  author       = {{Google}},
  howpublished = {\url{https://developer.chrome.com/docs/devtools}},
  year         = {2025},
  note         = {Accessed: 2025-09-09}
}

@article{Eisenstadt1997:Hairiest,
author = {Eisenstadt, Marc},
title = {My Hairiest Bug War Stories},
year = {1997},
issue_date = {April 1997},
publisher = {Association for Computing Machinery},
address = {New York, NY, USA},
volume = {40},
number = {4},
issn = {0001-0782},
url = {https://doi.org/10.1145/248448.248456},
doi = {10.1145/248448.248456},
journal = {Commun. ACM},
month = apr,
pages = {30--37},
numpages = {8}
}

@INPROCEEDINGS{Alaboudi2020:Hypotheses,
  author={Alaboudi, Abdulaziz and LaToza, Thomas D.},
  booktitle={2020 IEEE Symposium on Visual Languages and Human-Centric Computing (VL/HCC)}, 
  title={Using Hypotheses as a Debugging Aid}, 
  year={2020},
  volume={},
  number={},
  pages={1--9},
  doi={10.1109/VL/HCC50065.2020.9127273}}

@article{Vessey,
title = {Expertise in debugging computer programs: A process analysis},
journal = {International Journal of Man-Machine Studies},
volume = {23},
number = {5},
pages = {459--494},
year = {1985},
issn = {0020-7373},
doi = {https://doi.org/10.1016/S0020-7373(85)80054-7},
url = {https://www.sciencedirect.com/science/article/pii/S0020737385800547},
author = {Iris Vessey}
}

@article{stallman1988debugging,
  title={Debugging with GDB: The GNU Source-level Debugger},
  author={Stallman, Richard and Pesch, Roland and Shebs, Stan and others},
  journal={Free Software Foundation},
  volume={675},
  year={1988},
  ISBN={1-882114-77-9},
  location = {Boston, MA, USA},
  edition   = {9th.}
}

@INPROCEEDINGS{Layman2023:Debugging,
  author={Layman, Lucas and Diep, Madeline and Nagappan, Meiyappan and Singer, Janice and Deline, Robert and Venolia, Gina},
  booktitle={2013 ACM / IEEE International Symposium on Empirical Software Engineering and Measurement}, 
  title={Debugging Revisited: Toward Understanding the Debugging Needs of Contemporary Software Developers}, 
  year={2013},
  volume={},
  number={},
  pages={383--392},
  doi={10.1109/ESEM.2013.43}}

@INPROCEEDINGS{Beller2018:Dichotomy,
  author={Beller, Moritz and Spruit, Niels and Spinellis, Diomidis and Zaidman, Andy},
  booktitle={2018 IEEE/ACM 40th International Conference on Software Engineering (ICSE)}, 
  title={On the Dichotomy of Debugging Behavior Among Programmers}, 
  year={2018},
  volume={},
  number={},
  pages={572--583},
  doi={10.1145/3180155.3180175}}

@software{claude-sonnet,
  author       = {Anthropic},
  title        = {Claude Sonnet},
  year         = {2025},
  url          = {https://www.anthropic.com/claude},
  note         = {Large Language Model, Accessed: 2025-09-09},
}

@article{latoza2020explicit,
  title={Explicit programming strategies},
  author={LaToza, Thomas D and Arab, Maryam and Loksa, Dastyni and Ko, Amy J},
  journal={Empirical Software Engineering},
  volume={25},
  number={4},
  pages={2416--2449},
  year={2020},
  publisher={Springer},
  doi={https://doi.org/10.1007/s10664-020-09810-1}
}

@inproceedings{Chen2019:Towards,
author = {Li, Chen and Chan, Emily and Denny, Paul and Luxton-Reilly, Andrew and Tempero, Ewan},
title = {Towards a Framework for Teaching Debugging},
year = {2019},
isbn = {9781450366229},
publisher = {Association for Computing Machinery},
address = {New York, NY, USA},
url = {https://doi.org/10.1145/3286960.3286970},
doi = {10.1145/3286960.3286970},
booktitle = {Proceedings of the Twenty-First Australasian Computing Education Conference},
pages = {79--86},
numpages = {8},
location = {Sydney, NSW, Australia},
series = {ACE '19}
}

@article{McCauley01062008,
author = {Renée McCauley and Sue Fitzgerald and Gary Lewandowski and Laurie Murphy and Beth Simon and Lynda Thomas and Carol Zander},
title = {Debugging: a review of the literature from an educational perspective},
journal = {Computer Science Education},
volume = {18},
number = {2},
pages = {67--92},
year = {2008},
publisher = {Routledge},
doi = {10.1080/08993400802114581},
URL = { 
        https://doi.org/10.1080/08993400802114581
}
}

@article{ray_vibe_coding_2025,
title={A Review on Vibe Coding: Fundamentals, State-of-the-art, Challenges and Future Directions},
url={http://dx.doi.org/10.36227/techrxiv.174681482.27435614/v1},
DOI={10.36227/techrxiv.174681482.27435614/v1},
publisher={Institute of Electrical and Electronics Engineers (IEEE)},
author={Ray, Partha Pratim},
year={2025},
month=may }

@article{van_den_Haak01092003,
author = {Maaike van den Haak and Menno De Jong and Peter Jan Schellens},
title = {Retrospective vs. concurrent think-aloud protocols: Testing the usability of an online library catalogue},
journal = {Behaviour \& Information Technology},
volume = {22},
number = {5},
pages = {339--351},
year = {2003},
publisher = {Taylor \& Francis},
doi = {10.1080/0044929031000},
URL = { 
        https://doi.org/10.1080/0044929031000
},
}

@book{Ericsson_Simon_1993, title={Protocol Analysis: Verbal Reports as Data}, ISBN={978-0-262-27239-1}, url={https://doi.org/10.7551/mitpress/5657.001.0001}, DOI={10.7551/mitpress/5657.001.0001}, publisher={The MIT Press}, author={Ericsson, K. Anders and Simon, Herbert A.}, year={1993}, month=apr }

@INPROCEEDINGS{CokerFrameworkDebugging,
  author={Coker, Zack and Widder, David Gray and Le Goues, Claire and Bogart, Christopher and Sunshine, Joshua},
  booktitle={2019 IEEE International Conference on Software Maintenance and Evolution (ICSME)}, 
  title={A Qualitative Study on Framework Debugging}, 
  year={2019},
  volume={},
  number={},
  pages={568--579},
  doi={10.1109/ICSME.2019.00091}}

@inproceedings{RoehmSoftwareComprehension,
author = {Roehm, Tobias and Tiarks, Rebecca and Koschke, Rainer and Maalej, Walid},
title = {How do professional developers comprehend software?},
year = {2012},
isbn = {9781467310673},
publisher = {IEEE Press},
booktitle = {Proceedings of the 34th International Conference on Software Engineering},
pages = {255--265},
numpages = {11},
location = {Zurich, Switzerland},
series = {ICSE '12}
}

@inbook{computing_ed_today_2019, 
place={Cambridge}, 
series={Cambridge Handbooks in Psychology}, 
title={Computing Education Research Today}, 
booktitle={The Cambridge Handbook of Computing Education Research}, 
publisher={Cambridge University Press}, 
author={Fincher, Sally A. and Tenenberg, Josh and Dorn, Brian and Hundhausen, Christopher and McCartney, Robert and Murphy, Laurie}, 
editor={Fincher, Sally A. and Robins, Anthony V.Editors}, year={2019}, 
pages={40--55},
collection={Cambridge Handbooks in Psychology},
url={https://doi.org/10.1017/9781108654555.003}
}

@ARTICLE{Lawrance2013:Programmers,
  author={Lawrance, Joseph and Bogart, Christopher and Burnett, Margaret and Bellamy, Rachel and Rector, Kyle and Fleming, Scott D.},
  journal={IEEE Transactions on Software Engineering}, 
  title={How Programmers Debug, Revisited: An Information Foraging Theory Perspective}, 
  year={2013},
  volume={39},
  number={2},
  pages={197-215},
  doi={10.1109/TSE.2010.111}
}

@article{Brooks1983:Towards,
    title = {Towards a theory of the comprehension of computer programs},
    journal = {International Journal of Man-Machine Studies},
    volume = {18},
    number = {6},
    pages = {543-554},
    year = {1983},
    issn = {0020-7373},
    doi = {https://doi.org/10.1016/S0020-7373(83)80031-5},
    url = {https://www.sciencedirect.com/science/article/pii/S0020737383800315},
    author = {Ruven Brooks}
}

\end{document}